\DeclareUnicodeCharacter{2212}{-}
\documentclass{ijcaArticle}
\ijcaNumber{}
\ijcaYear{YYYY}
\ijcaMonth{Month}

\ijcaYear{}
\ijcaMonth{}
\begin{document}

\title{FeZnSb$_{2}$: A possible NiAs-type hexagonal superconductor  } 

\author{ 
   \large Phinifolo Cambalame$^{1,2}$ \\[-3pt]
   \normalsize $^{1}$ Centre for Physics of the University of Coimbra  \\[-3pt]
    \normalsize $^{2}$ Department of Physics, Eduardo Mondlane University \\[-3pt]
    \normalsize	phinifolo@fis.uc.pt \\[-3pt]
}

\terms{NiAs-type structure, partial disordered systems, superconductivity in Iron-based compounds}

\maketitle

\begin{abstract} 
 We present a first-principles investigation of electronic structure, lattice dynamics, and electron-phonon coupling of NiAs-type structure FeZnSb$_{2}$  and the isostructural parent compound FeSb within the framework of density functional theory. The calculation on partial disordered system FeZnSb$_{2}$ was performed in a fixed configuration. This hypothetical ordered structure is predicted to be superconducting. 
\end{abstract}

\section{Introduction}
The recent discovery of superconductivity in the NiAs-type structure high-entropy compound  M$_{1−x}$Pt$_{x}$Sb (M = equimolar Ru, Rh, Pd and Ir) \cite{Hirai2023} triggered our interest in the exploration of isostructural systems. Concurrently, the discovery of Fe-based pnictide and chalcogenide superconductors with high superconducting transition temperature ($T_c$) \cite{kamihara2008} has led to enormous excitement and opening of new perspective to develop a promising new platform to realize high-$T_c$ superconductors. Among these, FeZnSb$_{2}$ stands out - an unexplored Fe-based system that has a NiAs-type structure with disordered atomic positions of Fe and Zn (2\textit{a} Wyckoff position) while Sb atoms occupy the 2\textit{c} position  according to \cite{CHUMAK2000223} who first reported this system, derived from FeSb compound. To our knowledge, superconductivity has not been reported in an Fe-based compound with hexagonal structure, though the simple binary system FeSe, can crystallize in the superconducting phase $\ensuremath{\beta}$-FeSe (tetragonal) and $\ensuremath{\delta}$-FeSe (hexagonal) non-superconducting phase \cite{Zhang_2009, MENDOZA20101124, hechang2011}.  The occurrence of superconductivity within disordered NiAs-type structures \cite{Hirai2023} or enhancement of $T_c$ through disorder in other  crystal systems \cite{LEHMANN1981473} and the potential for ferroelectricity within distorted NiAs-type structures \cite{zhang2017PRB} render this material a confluence of features of great interest for both theoretical investigations and experimental studies. 
Disordered systems and partial ordered systems can also exhibit magnetic frustrated phases \cite{pakhira2017magnetic}, spin-glass \cite{You_2020} due to complex  exchange interactions between the magnetic atoms introduced by the atomic disorder \cite{Szlawska2011}. Partial ordered system generally have at least one ordered atomic position and remaining position randomly occupied. Fixed atomic configurations are relatively straightforward to handle and can yield data for comparison with experiments. Experimental parameters such as phonon vibrations and electronic structure can be calculated to infer the degree of disorder present in this system. Some systems with 50 $\%$ chemical scattering centers have already been observed to display superconductivity almost equal to pure elements \cite{MATTHIAS1956188}.

In our current investigation, we present an \textit{ab initio} prognosis regarding the superconductivity and stability of FeZnSb$_{2}$, under the assumption of fixed atomic positions for Zn and Fe. Our study entails the structural optimization of these compounds, juxtaposing the electronic structure and lattice dynamics of FeZnSb$_{2}$ with its parent compound, FeSb.

\section{Computational methods}
The DFT calculations were carried out on QUANTUM-ESPRESSO version 7.2 \cite{QE-2009, QE-2017, Giannozzi2020} using SSSP PBE Precision v1.2.0 pseudopotentials \cite{prandini2018precision}, except when mentioned otherwise. The crystal visualization was carried on Xcrysden software \cite{KOKALJ1999176}. The electronic band structure calculation were performed on relaxed structure,  sampling the Brillouin zone (BZ) with 16 $\times$ 16 $\times$ 12 k-points in the irreducible Brillouin zone (IBZ) with a 0.02 Ry smearing for BZ integration. The kinetic energy cutoff for wavefunctions was chosen to be 60 Ry and 420 Ry for charge density and potential for electronic structure and lattice dynamics calculations. As we are interested in superconductivity in the framework of DFT, other correlations effects beyond the DFT are not taken into account. Subsequently, to calculate the phonon dispersion curves and electron–phonon coupling we have employed density functional perturbation theory (DFPT), also within QUANTUM-ESPRESSO, using 20 $\times$ 20 $\times$ 16 electronic k-point grids and 5 $\times$ 5 $\times$ 4 phononic q-point grids. 

\section{Results}
\subsection{Mechanical and Lattice Parameters}
\begin{figure}
\centering
\begin{subfigure}{0.4\linewidth}
    \includegraphics[width=0.8\columnwidth, keepaspectratio]{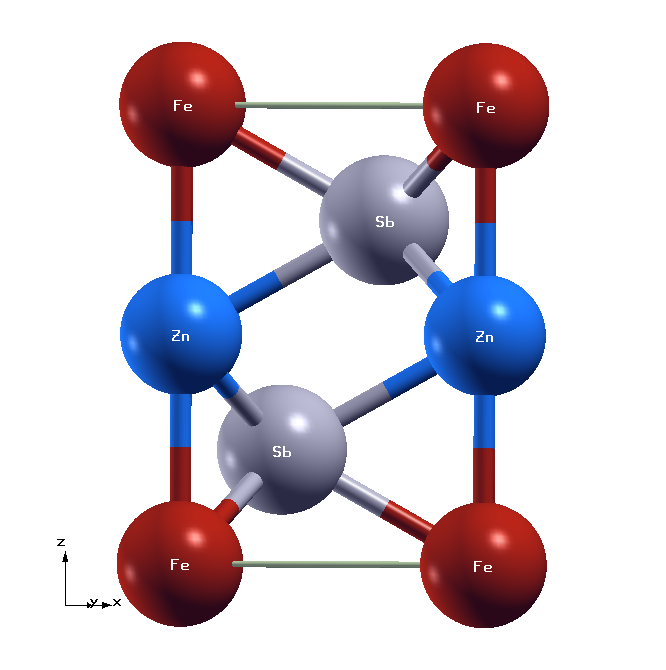}
\caption{}
\label{fig1a}
\end{subfigure}
\begin{subfigure}{0.4\linewidth}
    \includegraphics[width=0.8\columnwidth, keepaspectratio]{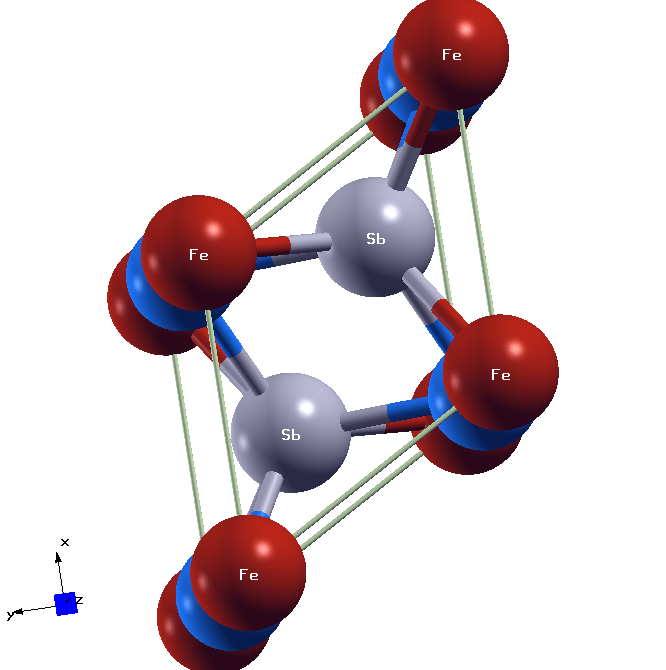}
\caption{}
\label{fig1b}
\end{subfigure}
\caption{FeZnSb$_2$. (a) View to xz-plane  and (b) xy-plane view.}
	\label{fig1}
\end{figure}

\indent
As mentioned earlier, FeZnSb$_{2}$ is a partially disordered system with NiAs-type crystal structure, while the Sb atoms occupy exclusively the  As- crystal site in NiAs, the Ni (2\textit{a}) site is now  occupied by Zn and Fe. The translational asymmetric unit cell representation gives two equivalent atomic site, that were atribute randomly to Zn and Fe atoms, resulting in the fixed or ordered configuration as depicted in Fig. \ref{fig1}, with the iron unit cell occupying the corner of the  unit cell , and the zinc atoms in the intermediate edge position between the Fe atoms. The inverse occupation (Zn and Fe interchanged) gives the same Fermi surface and phonon dispersions and electron-phonon coupling strength, as could be expected.  In FeSb, we have a similar picture, with the obvious difference being that Zn is now replaced by Fe atoms. 

\begin{table}[h]
\centering
\caption{Equilibrium crystal lattice parameters, \textit{a} and \textit{b}, Bulk modulus B and total energy \textit{E$_{min}$}.}
\begin{tabular}[t]{lccccc}
\hline
& \textit{a} &\textit{b} & B (GPa) &\textit{E$_{min}$  (Ry)}  & \\
\hline
FeSb & 3.99 & 4.97 & 140.4 & -1028.4026 \\
Exp. \cite{vasilev1982solid}& 4.1 & 5.14 & -- & --- \\\hline
FeZnSb$_{2}$& 4.19 &5.17  & 82.4 & -1160.70637 \\
Exp. \cite{CHUMAK2000223}       & 4.38 &  5.73  & -- & -- \\
\hline
\label{tab1}
\end{tabular}
\end{table}

\begin{figure}
\includegraphics[width=0.80\columnwidth, keepaspectratio]{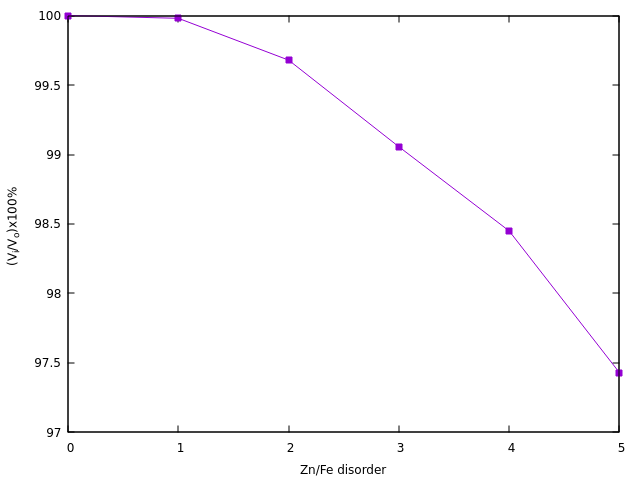}
\caption{Variation of ordered FeZnSb$_2$ with the Zn/Fe iter-site substitution}
\label{fig2}
\end{figure}

The optimization of the crystal structure from the experimental lattice parameters\cite{vasilev1982solid,CHUMAK2000223} provides good agreement between experimental and relaxed structure for FeSb but not in the FeZnSb$_{2}$ case. There's 10$\%$ disagreement of the \textit{c} lattice parameter, while for both FeSb lattice parameters this underestimation falls around $3\%$ (see  Table \ref{tab1}). To understand this uncommon underestimation  of  cell volume (and the \textit{c} lattice constant, in particularly) we performed a series of calculations with FeZnSb$_{2}$ supercells of 2 $\times$ 2 $\times$ 2 dimensions \cite{phonopy-phono3py-JPCM}, where the theoretical sites of Zn and Fe were gradually interchanged, the results are plotted at Fig. \ref{fig2}. Our calculations suggest that the increment of disorder in system, will shrink the unit cell volume. This result rules out disorder in the Zn and Fe positions as the primary cause for the abnormal underestimation of the lattice constant, thereby suggesting the potential involvement of other interactions, such as magnetism, in this discrepancy. We advance that the diffusion of Fe (or Zn) to the double tetrahedral interstitial positions within the primitive unit cell \cite{sladecek2001} constitutes the primary factor. This phenomenon has been previously documented in FeSb, NiSb, and related structures \cite{Hahnel1986}. Afterwards, we have performed magnetic calculations with different magnetic states and have come to the conclusion that the ground state of FeSb is ferromagnetic(FM), while for FeZnSb$_{2}$, the FM state and collinear antiferromagnetic(AFM) state are separated by a mere 2 mRy (calculation on 2 $\times$ 1 $\times$ 1 supercell). While the FM state predominates as the lowest-energy configuration,  this small difference suggest a more complex ground state and more detailed calculations is needed to elucidate further.

\subsection{Electronic Structure}
The calculated electronic band structure are shown in Fig. \ref{fig3}. There are relatively more bands crossing the Fermi level in FeSb  (Fig. \ref{fig3b}) than in FeZnSb$_{2}$ (Fig. \ref{fig3a}). There are multiple flat bands above the Fermi level along the $\Gamma$-M-K-$\Gamma$ in FeSb, and in FeZnSb$_{2}$ two degenerate flat bands along the L-H-K high-symmetry path are observed. The projected band structure calculations for FeZnSb$_{2}$ suggest a splitting into two sets of doubly degenerate states: one of the $d_{xz}$ and $d_{yz}$ character
and another of the $d_{{x^2} - {y^2}}$ and $d_{xy}$ character, along with a non-degenerate state derived from the  $d_{z^2}$ orbital. Unlike the FeZnSb$_{2}$ degeneracy picture, where we have a complete degeneracy along the displayed high-symmetry path (Fig. \ref{fig3a}), the set of $d_{xz}$ and $d_{yz}$ in FeSb have their degeneracy  lifted along the A-L-H path, the set of $d_{{x^2} - {y^2}}$ and $d_{xy}$ are partially degenerate only close to the Fermi level and significantly along M-K-$\Gamma$-A and lifted elsewhere. In FeZnSb$_{2}$, there's a significant hybridization between Sb ($p_{x}$ and $p_{y}$) with Fe ($d_{{x^2} - {y^2}}$). Hybridization  is marginally in FeSb.
\begin{center}
\centering
\begin{figure}
\begin{subfigure}{1.0\linewidth}
\includegraphics[width=0.80\columnwidth, keepaspectratio]{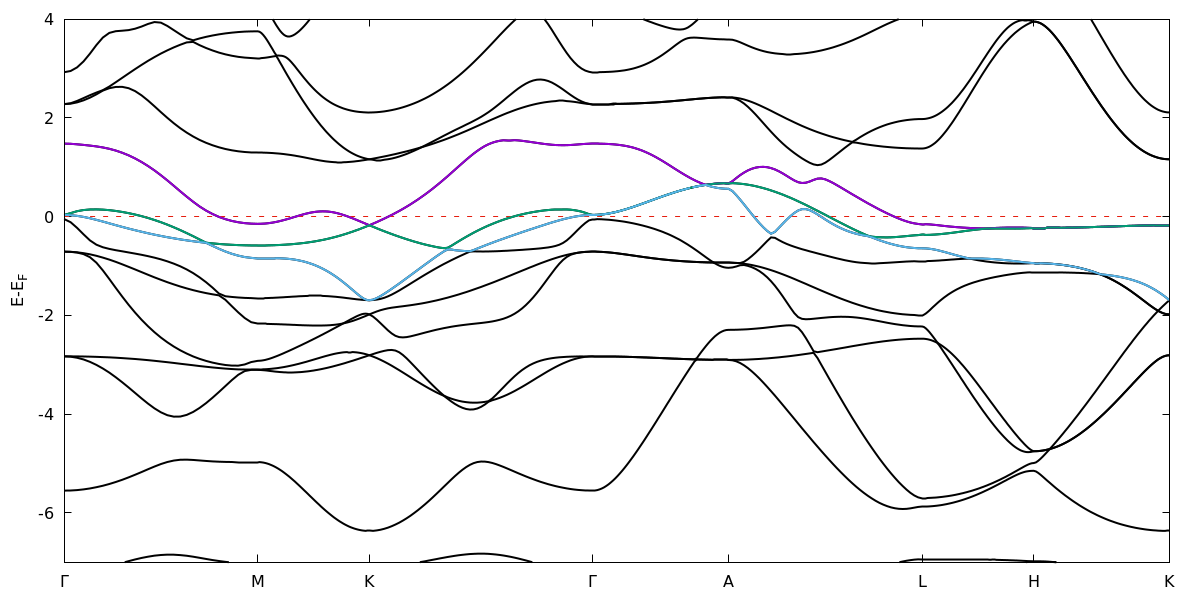}
\caption{}
\label{fig3a}
\end{subfigure}
\begin{subfigure}{1.0\linewidth}
\includegraphics[width=0.80\columnwidth, keepaspectratio]{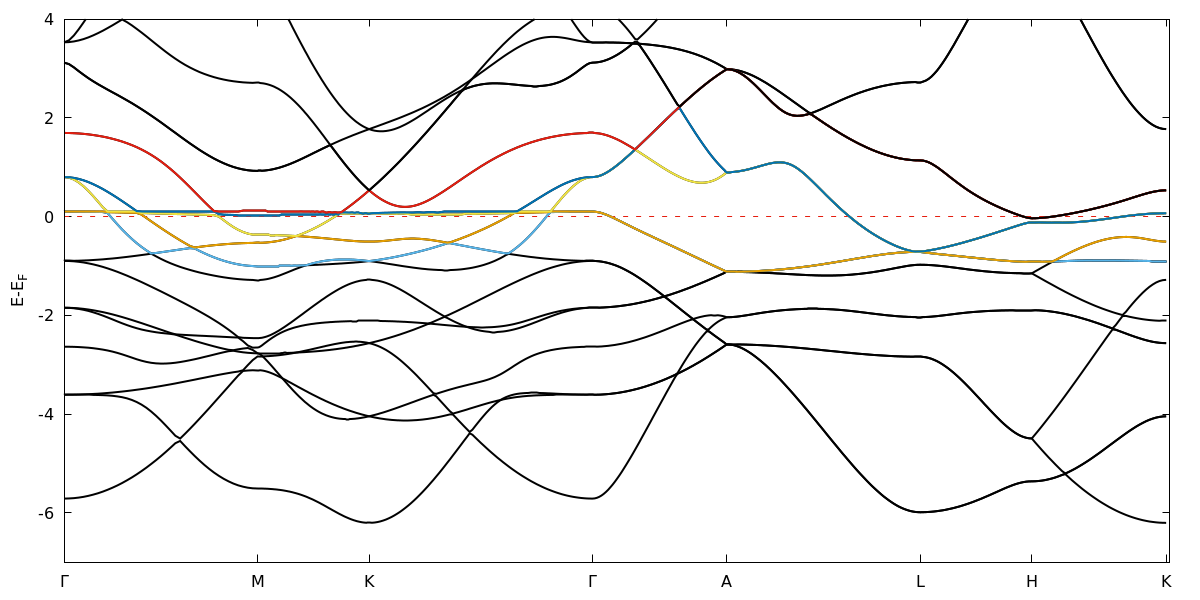}
\caption{}
\label{fig3b}
\end{subfigure}
\caption{Electronic structure of (a) FeZnSb$_2$  and (b) FeSb.}
	\label{fig3}
\end{figure}

\end{center}
\begin{figure}
\begin{subfigure}{1.0\linewidth}
\includegraphics[width=0.85\columnwidth, height=.4\textwidth]{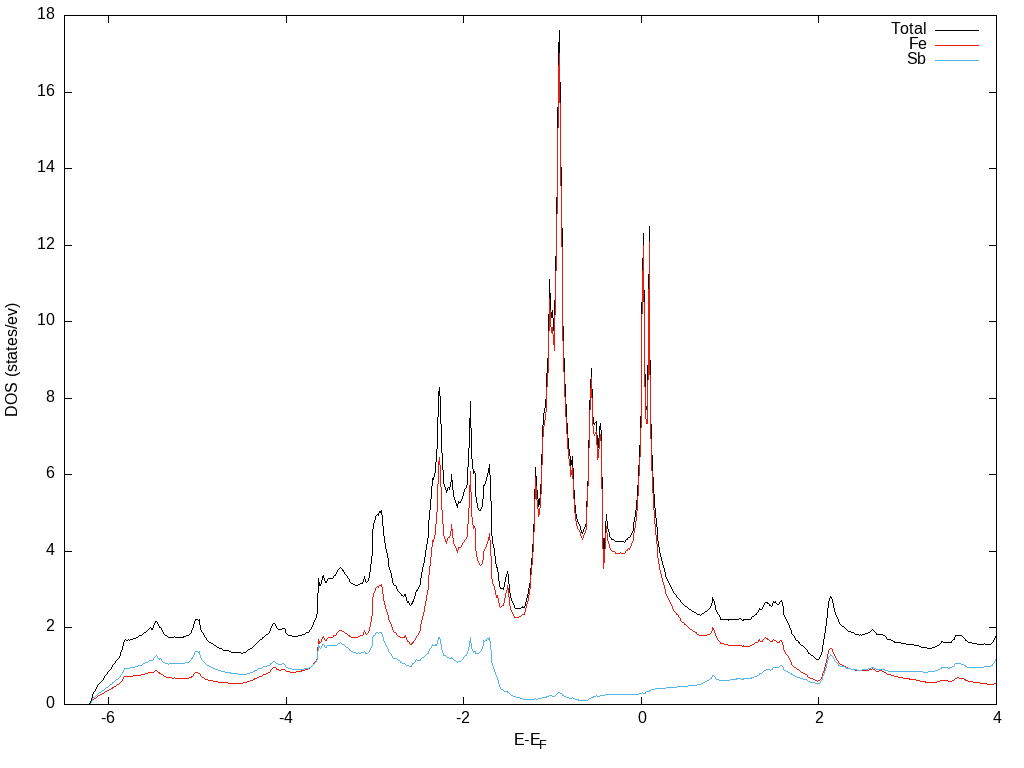}
\caption{}
\label{fig4a}
\end{subfigure}
\begin{subfigure}{1.0\linewidth}
\includegraphics[width=0.85\columnwidth, height=.4\textwidth]{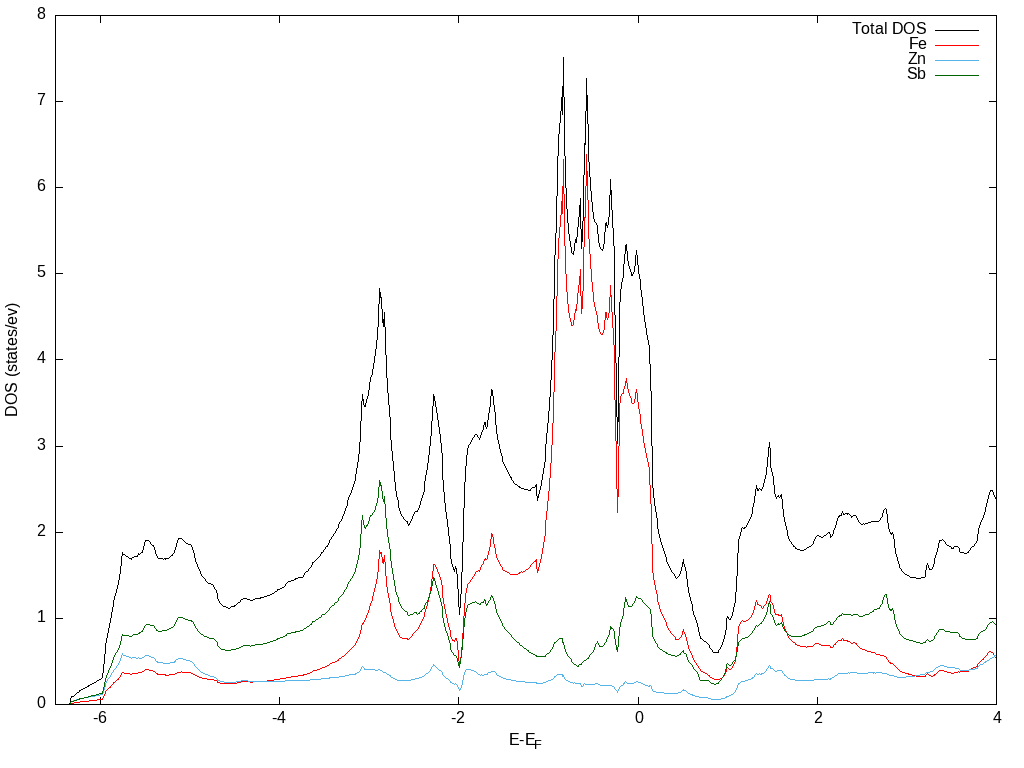}
\caption{}
\label{fig4b}
\end{subfigure}
\caption{Total and partial density of state (DOS) (a) for FeSb and (b) FeZnSb$_{2}$.}
	\label{fig4}
\end{figure}

The density of states at the Fermi level of these two materials exhibits a metallic behaviour (see Fig. \ref{fig4}). The contribution of Sb to the density of states is higher in FeZnSb$_{2}$ than in FeSb. This can be explained by the stronger covalent binding between Sb and Fe in the former compound. The contribution of Zn to the total DOS of FeZnSb is marginal. In both FeSb and FeZnSb$_{2}$ the highest DOS is observed along the L-H-K near the Fermi Level.

\subsubsection{Fermi surfaces}
\begin{figure}
\centering
\begin{subfigure}{0.4\linewidth}
 \includegraphics[width=0.8\columnwidth, keepaspectratio]{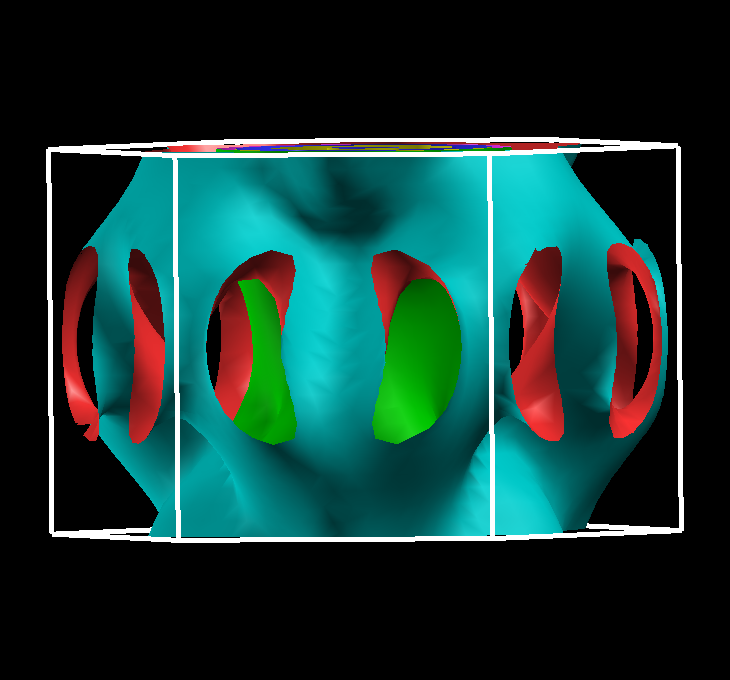}
\caption{}
\end{subfigure}
\begin{subfigure}{0.4\linewidth}
    \includegraphics[width=0.8\columnwidth, keepaspectratio]{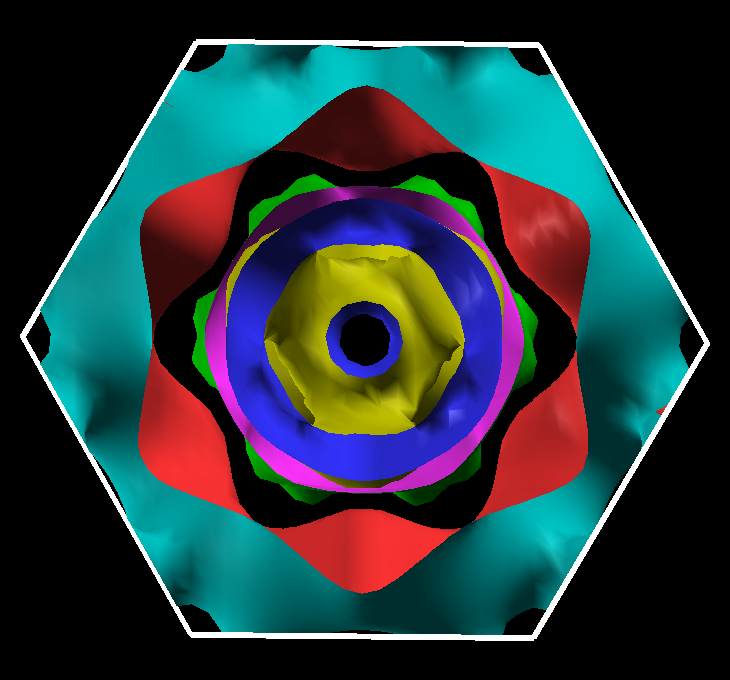}
\caption{}
\label{}
\end{subfigure}
\caption{Fermi surface of FeZnSb$_2$. (a) view to xz-plane, (b) xy-plane view.}
	\label{fig5a}
 \centering
\begin{subfigure}{0.4\linewidth}
    \includegraphics[width=0.8\columnwidth, keepaspectratio]{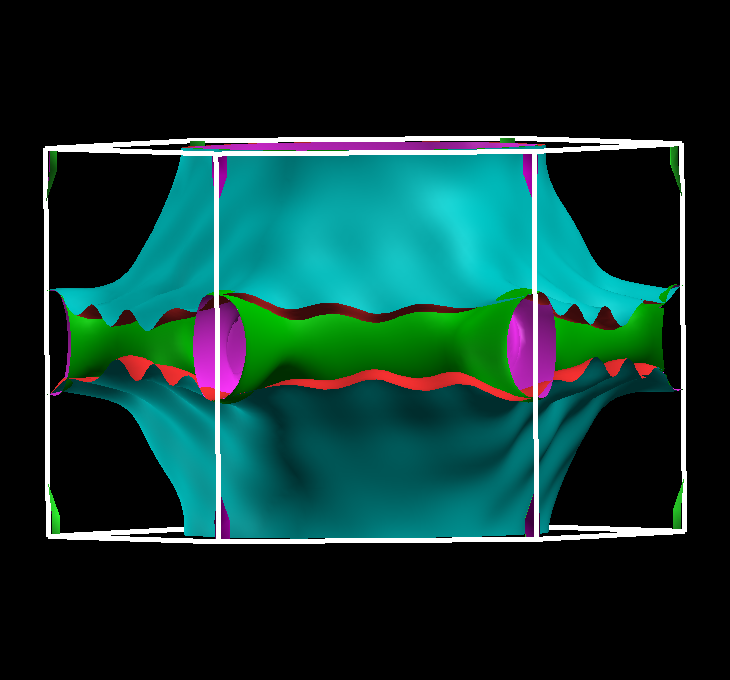}
\caption{}
\label{}
\end{subfigure}
\begin{subfigure}{0.4\linewidth}
    \includegraphics[width=0.8\columnwidth, keepaspectratio]{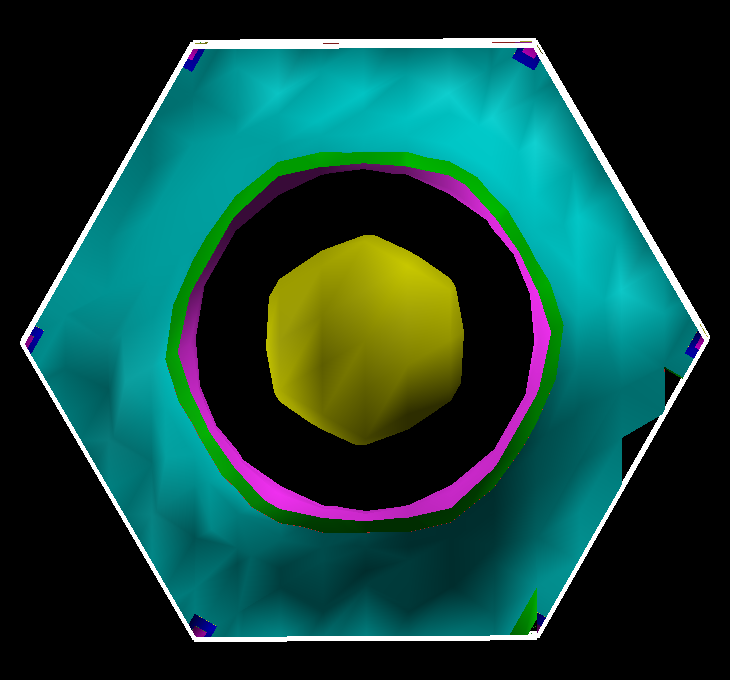}
\caption{}
\label{fig5b}
\end{subfigure}
\caption{Fermi surface FeSb. (a) view to xz-plane, (b) xy-plane view.}
	\label{fig5}
\end{figure}

At the $\Gamma$ point, we observe the primary topological difference in the Fermi surface (FS) see Fig. \ref{fig5a} and \ref{fig5}. In FeSb, we identify two anisotropic and co-centered 2-D hole pocket, whereas in FeZnSb$_{2}$ a cylinder-like structure encloses a complex dumbbell-like formation. In these two systems, we observe along the boundaries a typical
metallic open FS extending over almost the entire BZ. 

\subsection{Phonons and electron-phonon coupling}
\begin{figure}
\begin{subfigure}{1.0\linewidth}
\includegraphics[width=0.95\columnwidth, keepaspectratio]{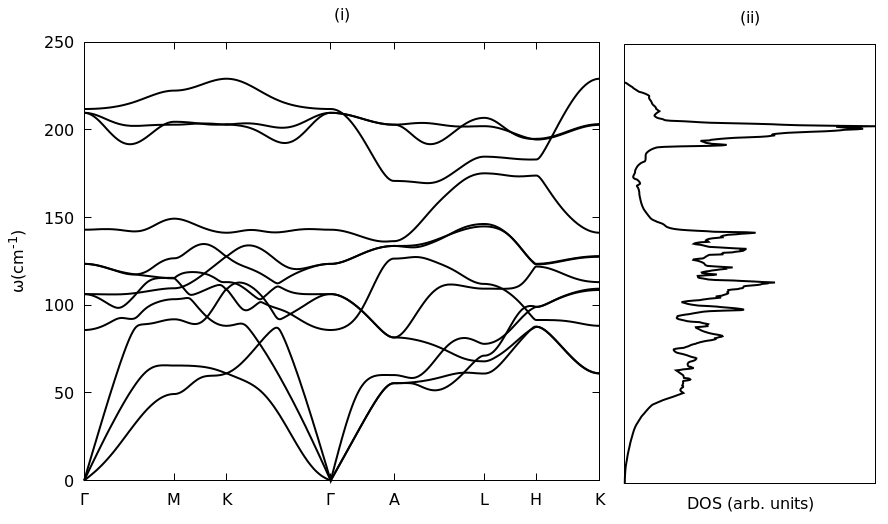}
\caption{}
\end{subfigure}
\begin{subfigure}{1.0\linewidth}
\includegraphics[width=0.95\columnwidth, keepaspectratio]{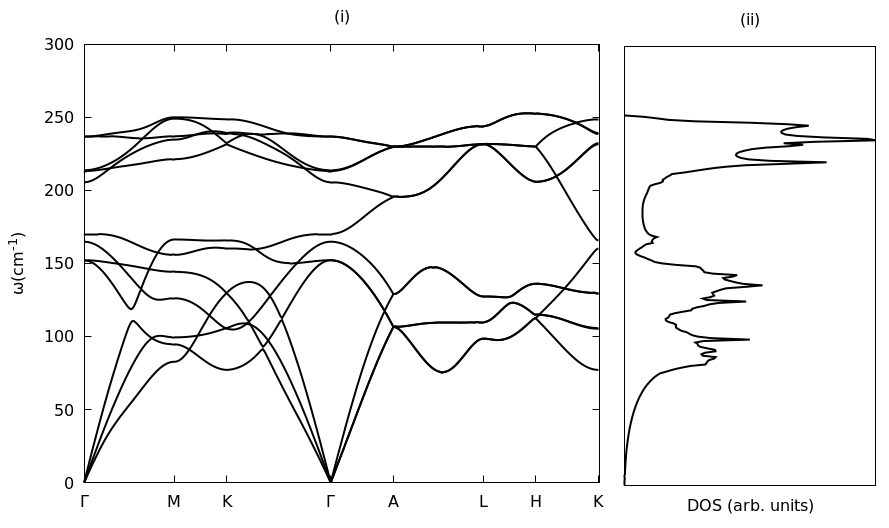}
\caption{}
\end{subfigure}
\caption{Phonon dispersion curves along high symmetry curves and the density of states (DOS) (a) for FeZnSb$_{2}$ and (b) FeSb .}
\label{fig6}
\end{figure}

The calculated  phonon dispersion curves are shown in Fig.\ref{fig6}. The absence of any imaginary or negative
vibrational modes suggests that both FeZnSb$_{2}$ and FeSb are dynamically stable systems within a hexagonal structure of NiAs-type. The origin of phonon branches is similar; the highest frequency branches result from the contribution of Fe and Zn in FeZnSb$_{2}$ and Fe in FeSb. The three (two) atomic species
are involved in the acoustic vibration modes of FeZnSb$_{2}$ (FeSb). This pattern of vibration explains the depression in the phonon DOS when transiting to high frequencies. Raman and infrared (IR) were calculated at $\Gamma$. The calculation on the FeZnSb$_{2}$ (FeSb) relaxed structure assigns the crystallographic symmetry to D$_{3d}$ (C$_{2v}$), which decomposes into the vibrational modes in ($\Gamma$) to the following irreducible representations:

\begin{equation}
    \Gamma = 3 A_{2u} + 3 E_{u} + E_{g} + A_{1g}
\end{equation}

\begin{table}[H]
\caption{The calculated frequencies of the Raman (R) and infrared (I) active modes of FeSb and FeZnSb$_2$. The modes are classified by the irreducible representations (irreps) according to which they transform.}
\begin{tabular}[t]{lccccr}

  \hline
       FeSb &  & &FeZnSb$_2$ &  \\
 \hline
    $\omega$ cm$^{-1}$&  (irrep)  & &$\omega$ cm$^{-1}$&(irrep) & \\
 \hline
          8.1 & ($B_{1}$)       & I+R & 11.3 &  ($A_{2u}$) & I \\
         152.1  & ($A_{1}, B_{2}$) & I+R & 17.7  &($E_{u}$)  & I \\
         164.8  & ($A_{2}$)       & R   & 88.2 & ($A_{2u}$) & I \\
         170.5  & ($B_{1}$)       & I+R & 108.1 & ($E_{u}$)  & I \\
         205.9  & ($B_{2}$)       & I+R & 125.3 & ($E_{g}$)  & R \\
         213.3  & ($A_{2}$)       & R   & 143.1 & ($A_{1g}$) & R \\
         213.7  & ($B_{1}$)       & I+R & 209.3 & ($E_{u}$)  & I \\
         236.8  & ($A_{1}$)       & I+R & 212.7 & ($A_{2u}$) & I \\

   \hline
   \label{tab2}
\end{tabular}
\end{table}



The calculated Raman and infrared active modes can be validated through experimentation, enabling inference about the disorder of the system and its crystallographic symmetry, particularly in the case of  FeZnSb$_{2}$. The $E_{g}$ Raman mode exhibits degeneracy. Both Raman modes are attributed exclusively to Sb vibrations in FeZnSb$_{2}$, whereas in FeSb, one Raman mode (164.8 cm$^{-1}$) originates from Sb and 213.13 cm$^{-1}$ from Fe. The presence of flat bands in this material may suggest low thermal conductivity and potential thermoelectric capacity \cite{fan2017understanding}.
\subsubsection{Electron-phonon coupling and superconductivity}

\begin{figure}
\begin{subfigure}{0.5\linewidth}
\includegraphics[width=0.9\columnwidth, keepaspectratio]{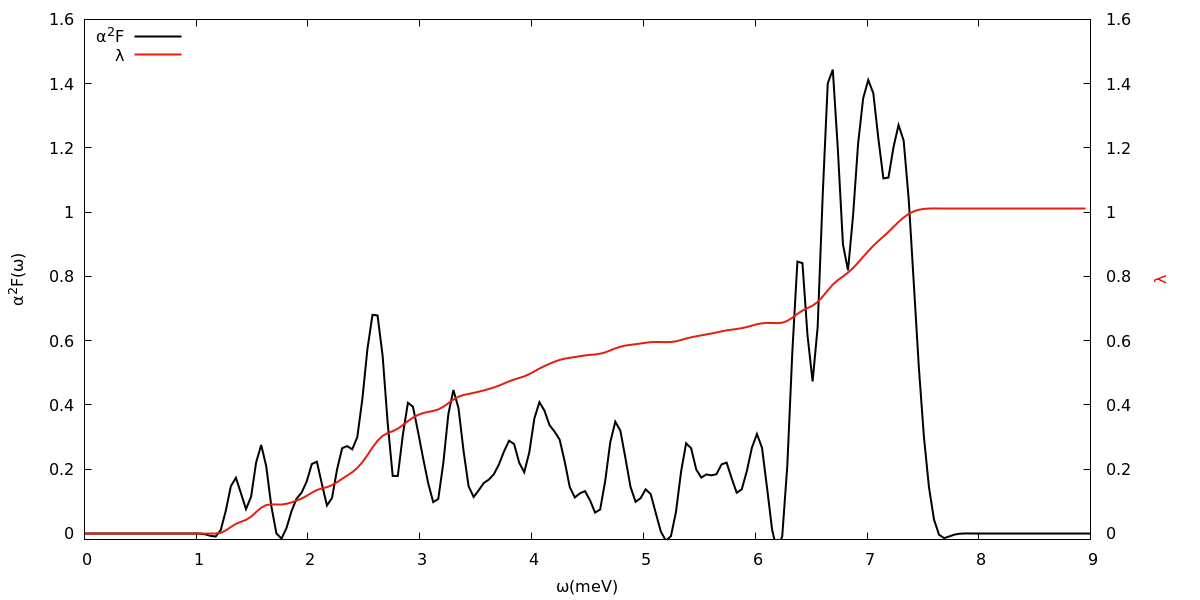}
\caption{}
\label{fig7b}
\end{subfigure}
\begin{subfigure}{0.5\linewidth}
\includegraphics[width=0.9\columnwidth, keepaspectratio]{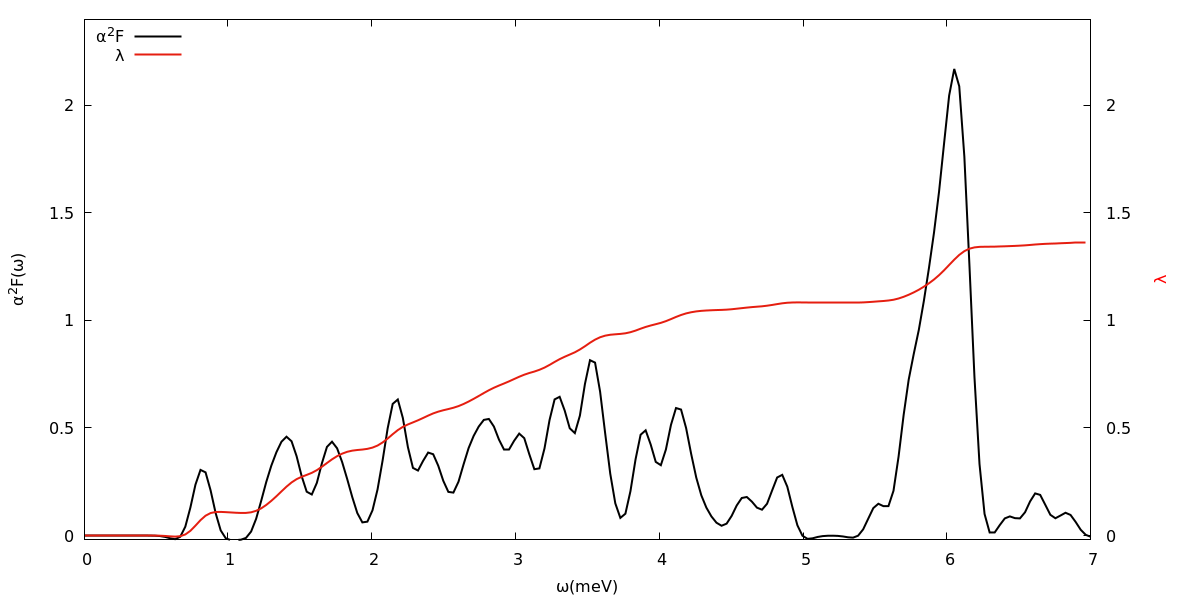}
\caption{}
\label{fig7c}
\end{subfigure}
\caption{ Electron-phonon spectral function $\alpha^{2}F(\omega)$ and cumulative electron-phonon coupling strength $ \lambda $ for (a) FeSb and  (b) FeZnSb$_{2}$}
	\label{fig7}
\end{figure}

The frequency dependence of the Eliashberg spectral function $\alpha^{2}F(\omega)$  and the average $\lambda$ value for FeSb and FeZnSb$_{2}$ are presented in Table \ref{tab3} and Fig. \ref{fig7}. We have also calculated the relevant parameters for the
description of the superconducting properties within the
Eliashberg formalism, namely the logarithmically averaged phonon frequency $\omega_{log}$ and the superconducting transition temperature $T_{c}$, calculated by the McMillan \cite{McMillan1968} formula modified  by Allen-Dynes \cite{Allen_Dynes1975} Eq. \ref{eq2}, presented in the table \ref{tab3}.
\begin{equation}
    T_c = \frac{\omega_{log}}{1.2}exp \left( -\frac{1.04(1 + \lambda )}{\lambda - \mu^{*}(1 + 0.62\lambda)} \right)
    \label{eq2}
\end{equation}

\begin{table}
\centering
\caption{The calculated DOS at E$_F$, N(0), the average electron–phonon
coupling constant $\lambda$, the
logarithmically averaged  $\omega_{log}$ phonon frequencies and \textit{T}$_{c}$  calculated from modified McMillan formula eqn. \ref{eq2}}
\begin{tabular}[t]{lccccc}
\hline
& \textit{N(0)} &$ \lambda $ &  $\omega_{log} [K]$  &\textit{T$_{c} [K]$  ($\mu = 0.1$)}  & \\
\hline
FeSb        & 39.471550 & 0.33 & 191.886 & 0.217 \\
FeZnSb$_{2}$& 21.530679 & 0.59 & 139.067 & 3  \\
\hline
\label{tab3}
\end{tabular}
\end{table}

To validate the accuracy of our findings, we conducted a fresh computation employing the projector augmented-wave (PAW) pseudopotentials \cite{Jollet2014} version 2.0.1, yielding closely aligned results. Specifically, we observed an average T$_{c}$ of 3.7 K and an electron-phonon coupling strength of 0.66.
\subsection{Discussion and conclusions}

\indent
In this study, we demonstrate the stability of synthesized compounds FeSb and FeZnSb$_{2}$ with the NiAs structure. The simulated disorder in FeZnSb$_{2}$ reveals that the lattice parameter will shrink with disorder, and the substituion of Fe by Zn increases the relevance of Sb in the DOS next to the Fermi level. If the phonon mediated pairing mechanism for superconductivity is assumed for these systems, superconducting transition temperature is expected to be significant in the unexplored system FeZnSb$_{2}$. While disorder within the system may negatively impact superconductivity, the potential occurrence of superconductivity in this Fe-based compound offers a novel platform for material manipulation and comparison with unconventional superconductors, particularly Fe-based layered compounds. The findings of this theoretical study serve as strong motivation for the synthesis and experimental exploration of this compound, as well as for employing more sophisticated methodologies in theoretical calculations. This includes the introduction of supercells and disorder to better align with the realistic characteristics of the FeZnSb$_{2}$ system.

\begin{acknowledgments}
This work was produced with the support of CFISUC computing resources, cluster Adamastor.
\end{acknowledgments}

%
%
%
%

\bibliographystyle{ijcaArticle}
\bibliography{literature}

\end{document}